\documentclass[prd,amssymb,11pt]{revtex4}
\usepackage{epsfig}
\newcommand{\beq}{\begin{equation}}
\newcommand{\eeq}{\end{equation}}
\newcommand{\bqa}{\begin{eqnarray}}
\newcommand{\eqa}{\end{eqnarray}}
\newcommand{\fr}{\frac}

\begin{document}
\title{Global symmetries: matter from geometry, hoop conjecture, and cosmic censorship}
\author{S\'{e}rgio M. C. V. Gon\c{c}alves\footnote{E-mail address: {\tt sergio.goncalves@yale.edu}}}
\affiliation{Department of Physics, Yale University, New Haven, Connecticut 06511}
\date{\today}
\begin{abstract}
We show that four-dimensional Lorentzian metrics admitting a global spacelike Lie group of isometries, $G_{1}={\mathbb R}$, which obey the Einstein equations for vacuum and certain types of matter, cannot contain apparent horizons. The assumed global isometry allows for the dimensional reduction of the $(3+1)$ system to a $(2+1)$ picture, wherein the four-dimensional metric fields act formally as matter fields. A theorem by Ida allows one to check for the absence of apparent horizons in the dimensionally reduced spacetime, with the four-dimensional results following from the topological product nature of the corresponding manifold. We argue that the absence of apparent horizons in spacetimes with translational symmetry constitutes strong evidence for the validity of the hoop conjecture, and also hints at possible (albeit arguably unlikely) generic violations of strong cosmic censorship.
\end{abstract}
\maketitle

\newpage

%
%

The notion of ``horizon'' plays a crucial role in classical general relativity: event horizons are the defining property of black holes, apparent horizons are numerical relativity's black hole operational definition, and particle horizons characterize many cosmological models and are responsible for the formation of topological defects. In addition, two of the outstanding issues in the classical theory---the cosmic censorship and hoop conjectures---rely crucially on the notion(s) of event and/or apparent horizon. In this essay, we shall focus on the latter, since it not only provides a practical definition of black hole, but it also implies the existence of an event horizon, provided weak cosmic censorship and certain asymptotic conditions are satisfied~\cite{hawking&ellis73}. 

While the relation between ``mass'' and horizon has been largely explored in the literature, that between geometry and horizon has received comparatively less attention. A notable exception is the hoop conjecture, which uses the combined roles of geometry and mass to put forward a (loosely formulated) necessary and sufficient condition for horizon formation~\cite{thorne72}: {\em ``Horizons form when and only when a mass $M$ gets compacted into a region whose circumference in {\bf every} direction is $C\lesssim4\pi M$''}. Despite inherent ambiguities in the definitions of horizon, mass, and circumference, no credible counter-example appears to exist~\cite{nocounterhc}. However, none of these answers the fundamental question: {\em Why is it that mass needs to be compacted in all three spatial directions for a horizon to form?}

Thus motivated, we consider here a large class of spacetimes which cannot, by construction, admit spatially bounded mass configurations, and show that they cannot contain apparent horizons (outer marginally trapped surfaces which are the outer boundary of a trapped region). Specifically, we consider four-dimensional spacetimes $(M,{\mathbf g})$ with the single assumption that they admit a global spacelike isometry group $G_{1}={\mathbb R}$ acting on a three-manifold ${\mathcal M}\approx{\mathbb R}\times\Sigma$, where $\Sigma$ is a spacelike two-surface of arbitrary topology. The two key ingredients in the analysis are (i) the dimensional reduction of the $(3+1)$ problem to $(2+1)$ form, enabled by the global isometry, and (ii) a theorem by Ida~\cite{ida00}, which provides a sufficient condition for the absence of apparent horizons in three-dimensional spacetimes. The program is then to take a $G_{1}$-symmetric four-dimensional spacetime, perform the dimensional reduction, test for the absence of apparent horizons in the dimensionally reduced picture, and then go back to the full $(3+1)$ system and use its topological product structure to infer the topology of the apparent horizons that cannot exist therein. In what follows, we outline the vacuum case for simplicity, and then discuss generalizations to include matter fields.

{\bf Proposition 1} (Dimensional Reduction): {\em Let $(M,\mathbf{g})$ be a $(3+1)$ spacetime obeying the vacuum Einstein equations ${\mathbf G}=0$. If ${\mathbf g}$ admits $G_{1}={\mathbb R}$ as a global spacelike isometry Lie group then the following is true: (i) The equations ${\mathbf G}=0$ are equivalent to $^{(3)}{\mathbf G}=\,^{(3)}{\mathbf T}_{\rm r}$, where $^{(3)}{\mathbf G}$ is the Einstein tensor in the quotient space ${\mathcal M}=M/{\mathbb R}$, and $^{(3)}{\mathbf T}_{\rm r}$ is the stress-energy tensor of a massless scalar field coupled to an electromagnetic-like vector potential field, defined on ${\mathcal M}$; (ii) $^{(3)}{\mathbf T}_{\rm r}$ obeys the dominant energy condition.}

In what follows, we adopt Moncrief's Hamiltonian reduction approach to vacuum gravity with one spacelike Killing vector field (KVF)~\cite{moncrief86}, and take the isometry Lie group to be $G_{1}={\mathbb R}$. Let the coordinates in $M$ be $\{x^{3},x^{i}; i=0,1,2\}$, and let the KVF be $\partial_{x^{3}}$, whose space of orbits under $G_{1}$ actions induces a three-manifold ${\mathcal M}=M/{\mathbb R}$ . The four-metric in $M$ can then be written as
\beq
ds^{2}=e^{-2\phi}\gamma_{ij}dx^{i}dx^{j}+e^{2\phi}(dx^{3}+\beta_{a}dx^{a}+\beta_{0}dt)^{2}, \label{msplit}
\eeq
where $|\partial_{x^{3}}|\equiv e^{\phi}$, and the induced Lorentzian metric in ${\mathcal M}\approx{\mathbb R}\times\Sigma$ admits the ADM decomposition
\beq
\gamma_{ij}dx^{i}dx^{j}=-\tilde{N}dt^{2}+\tilde{\sigma}_{ab}(dx^{a}+\tilde{N}^{a}dt)(dx^{b}+\tilde{N}^{b}dt), 
\eeq
where the indices $(a,b,c,...)$ refer to two-dimensional quantities, denoted by a tilde, defined on $\Sigma$. Introducing momenta $(\tilde{p}, \tilde{e}^{a}, \tilde{\pi}^{ab})$ conjugate to $(\phi, \tilde{\beta}_{a}, \tilde{\sigma}_{ab})$, the Einstein-Hilbert action is
\beq
S=\int_{\mathcal M} dt d^{2}x (\tilde{\pi}^{ab}\tilde{\sigma}_{ab,t}+\tilde{e}^{a}\tilde{\beta}_{a,t}+\tilde{p}\phi_{,t}-\tilde{N}\tilde{\mathcal H}-\tilde{N}^{a}\tilde{\mathcal H}_{a}-\beta_{0}\tilde{e}^{a}_{,a}), \label{eha}
\eeq
where the canonical Hamiltonian scalar and momentum vector densities are, respectively,
\bqa
\tilde{\mathcal H}&=&\fr{1}{\sqrt{\tilde{\sigma}}}[\tilde{\pi}^{ab}\tilde{\pi}_{ab}-(\tilde{\pi}^{a}_{a})^{2}+\fr{1}{8}\tilde{p}^{2}+\fr{1}{2}e^{-\phi}\tilde{\sigma}_{ab}\tilde{e}^{a}\tilde{e}^{b}] \nonumber \\
&&+\sqrt{\tilde{\sigma}}\{-^{(2)}\!\tilde{R}+2\tilde{\sigma}^{ab}\phi_{,a}\phi_{,b}+e^{4\phi}\tilde{\sigma}^{ac}\tilde{\sigma}^{bd}\tilde{\beta}_{[a,b]}\tilde{\beta}_{[c,d]}\}, \label{hcs} \\
\tilde{\mathcal H}_{a}&=&-2\tilde{\nabla}_{b}\tilde{\pi}^{b}_{a}+\tilde{p}\phi_{,a}+2\tilde{e}^{b}\tilde{\beta}_{[b,a]}. \label{mcs}
\eqa
The constraint equations for the action $S$ are
\beq
\tilde{\mathcal H}=0, \;\;\;\; \tilde{\mathcal H}_{a}=0, \;\;\;\; \tilde{e}^{a}_{,a}=0,
\eeq
and are equivalent to the four-dimensional constraints, restricted to the assumed symmetry class. Using Eqs. (\ref{hcs})-(\ref{mcs}), the action (\ref{eha}) can be written as:
\beq
S=S_{\rm G}+S_{\rm M}=\int_{\mathcal M} dt d^{2}x(\tilde{\pi}^{ab}\tilde{\sigma}_{ab,t}-\tilde{N}\tilde{H}-\tilde{N}^{a}\tilde{J}_{a})-\int_{\mathcal M} \sqrt{|\gamma|} d^{3}x \{2\phi_{,i}\phi^{,i}+\fr{e^{4\phi}}{2}\Psi_{j}^{\;\,k}\Psi^{j}_{\;k}\},
\eeq
with
\beq
\tilde{H}=\fr{1}{\sqrt{\tilde{\sigma}}}[\tilde{\pi}^{ab}\tilde{\pi}_{ab}-(\tilde{\pi}^{a}_{a})^{2}]-\sqrt{\tilde{\sigma}}\,^{(2)}\!\tilde{R}, \;\;\;\;\;\; \tilde{J}_{a}=-2\tilde{\nabla}_{b}\tilde{\pi}^{b}_{a}, \;\;\;\;\;\; \Psi_{ij}=2\beta_{[j,i]},
\eeq
where $S_{\rm G}$ is just the canonical action for pure $(2+1)$ gravity, and $S_{\rm M}$ is the action associated with ``matter'' fields $\phi$ and $\tilde{\beta}_{a}$. The canonical stress-energy tensor derived from $S_{\rm M}$ is:
\beq
T_{ij}:=-\fr{1}{\sqrt{|\gamma|}}\fr{\delta S_{\rm M}}{\delta \gamma^{ij}}=\phi_{,i}\phi_{,j}-\fr{1}{2}\gamma_{ij}\phi_{,k}\phi^{,k}+\fr{e^{4\phi}}{2}(\Psi_{i}^{\;k}\Psi_{jk}-\fr{1}{4}\gamma_{ij}\Psi_{m}^{\;\;n}\Psi^{m}_{\;\;n}),
\eeq
which has the form of a massless scalar field coupled to an ``electromagnetic'' field. Hamilton's equations guarantee that $T_{ij}$ is divergence-free, and it is easy to show that $T_{ij}$ also obeys the dominant energy condition (DEC).

{\bf Proposition 2} (Ida's Theorem): {\em Let $({\mathcal M},\mathbf{\gamma})$ be a $(2+1)$-dimensional Lorentzian spacetime satisfying the Einstein equations ${\mathbf G}( \gamma)={\mathbf T}$. If ${\mathbf T}$ obeys the dominant energy condition, then there are no apparent horizons in $({\mathcal M},\mathbf{\gamma})$.}

The idea of the proof consists in showing that, if an apparent horizon ${\mathcal A}$ exists {\em and} the dominant energy condition is satisfied, then one could deform ${\mathcal A}$ outward, so as to produce a new closed surface $\hat{\mathcal A}$ just outside ${\mathcal A}$, which is contained in a trapped region, thereby contradicting the ansatz that the former is the outer boundary of a compact trapped region. Ida's statement of the theorem~\cite{ida00} includes a positive cosmological constant, $\Lambda>0$, but the same result can be proved (with minor technical differences) for $\Lambda=0$~\cite{goncalves03a}.

{\bf Theorem 1} (Vacuum): {\em Let $(M,\mathbf{g})$ be a four-dimensional Lorentzian spacetime obeying the vacuum Einstein equations ${\mathbf G}(\mathbf{g})=0$. If $(M,\mathbf{g})$ admits a global spacelike $G_{1}={\mathbb R}$ Lie group of isometries, then it cannot contain apparent horizons.}

From Propositions 1 and 2 it follows that there are no apparent horizons in $({\mathcal M}=M/{\mathbb R},\gamma)$, which arises from dimensional reduction of vacuum $(3+1)$ gravity under $G_{1}$ actions. Since $M$ is topologically the Cartesian product ${\mathcal M}\times{\mathbb R}$, it follows that the absence of topological $S^{1}$ apparent horizons in $({\mathcal M},\gamma)$ implies that there are no apparent horizons homeomorphic to $S^{1}\times{\mathbb R}$ in $(M,{\mathbf g})$. Conversely, suppose that there is some three-dimensional trapped region ${\mathcal R}$, homeomorphic to $S^{2}$ or $S^{1}\times{\mathbb R}$, satisfying ${\mathcal R}\subset {\mathcal U}$, where ${\mathcal U}$ is a spacelike three-surface in $M$. Since the $G_{1}$ group orbits are spacelike, one can always find a two-dimensional foliation $\Sigma_{*}$ of ${\mathcal U}$ whose intersection with the outer boundary of the trapped region is: (i) a closed curve, i.e., $\exists\,\Sigma_{*}\approx{\mathbb R}^{2}: {\mathcal U}={\mathbb R}\times\Sigma_{*}$ and $\Sigma_{*}\cap\dot{\mathcal R}={\mathcal A}\approx S^{1}$, and (ii) ${\mathcal A}$ is an apparent horizon in $({\mathcal M},\gamma)$. But this contradicts Proposition 2, and thus apparent horizons cannot exist in $(M,{\mathbf g})$~\cite{goncalves03a}.

{\bf Theorem 2} (Matter): {\em Let $(M,\mathbf{g})$ be a four-dimensional Lorentzian spacetime obeying the Einstein equations ${\mathbf G}(\mathbf{g})={\mathbf T}$, and let $G_{1}={\mathbb R}$ be a global spacelike group of isometries in $(M,\mathbf{g})$. If the stress-energy tensor of the dimensionally reduced space $M/{\mathbb R}$ obeys the dominant energy condition, then $(M,{\mathbf g})$ cannot contain apparent horizons.}

This result follows from Proposition 2, together with the method of proof for Theorem 1. The Einstein-Maxwell case is one example that obeys this theorem~\cite{goncalves03b}. The ${\mathbb R}$-reduction of the system yields $(2+1)$ gravity coupled to four scalar fields, which define a wave map from the reduced spacetime to a target space $({\mathbb R}^{4},{\mathbf h})$, where ${\mathbf h}$ is a Riemaniann metric~\cite{moncrief90}. The canonical stress-energy tensor associated with wave maps (of {\em any} dimension) obeys the DEC, so by the arguments outlined above it follows that $(3+1)$ electrovacuum spacetimes with spacelike translational invariance cannot contain apparent horizons~\cite{goncalves03b,goncalves03c}.

These results deal with large classes of spacetimes which, by construction, can only admit confinement of mass along two spacelike directions: such spacetimes cannot develop apparent horizons. This provides very strong evidence towards the ``only if'' part of the hoop conjecture (if a horizon exists then mass is sufficiently compacted along all three spatial directions), by proving its converse: if mass is not compacted along all three spatial directions then apparent horizons cannot exist. The theorems rule out cases that would be blatant violations of the conjecture: if an infinite topological cylinder $S^{1}\times{\mathbb R}$ could be outer marginally trapped this would be an example of mass being only compacted along two spatial directions with an apparent horizon being nevertheless present. We point out that this no-horizon property of gravity with a translational spacelike KVF is not a mere geometrical artifact; rather, it is a genuine feature of the theory, enforced by the field equations. We further note that the inclusion of a {\em positive} cosmological constant leaves our conclusions unchanged, since it preserves the DEC. 

Finally, this result also has potential implications for strong cosmic censorship for gravity coupled to well-behaved (in the DEC sense) matter or vacuum, with the chosen KVF. The absence, to date, of large data global hyperbolicity results for vacuum or matter-coupled Einstein equations with one spacelike KVF leaves open the possibility for nonspacelike singularities, which would spoil global hyperbolicity~\footnote{This is possible to the extent that its converse has not yet been proven. However, existing long-term, large data results for vacuum {\em cylindrical} spacetimes (which have two KVF's)~\cite{bcm97}, together with several local (i.e., ``short time'') existence results for spacetimes with a single KVF, strongly suggest that global hyperbolicity should hold for vacuum spacetimes with one-dimensional spacelike Lie group(s) of isometries. I thank Vince Moncrief for enlightening discussions concerning this point.} . Any such singularities would be at least locally naked, since every nonspacelike geodesic emanating from the singularity would be untrapped, thus in clear violation of the strong cosmic censorship conjecture. The implications for weak cosmic censorship are less clear, since the former requires an event horizon, the definition of which for non-asymptotically-flat spacetimes is still lacking~\footnote{Such definition requires, among other things, the knowledge of radiative fall-off rates at asymptotic infinity. One of the main problems is the difficulty in characterizing the asymptotics of such spacetimes; even in the particular case of cylindrical spacetimes with polarized vacuum, different families of static cylinders admit different definitions of ${\mathcal J}^{+}$~\cite{ashtekar97}.}. Although it seems likely that event horizons (defined in a suitable sense) cannot exist either---the no-horizon proof is {\em independent} of a particular spacelike foliation choice, and it is difficult to imagine a spacetime that is free of closed trapped regions for {\em every} possible $(3+1)$ slicing and yet contains an event horizon---because of ambiguities with the definition of the former we shall not pursue this discussion here.

\end{document}